\documentclass[useAMS,usenatbib,a4]{mn2e}
\usepackage{amsmath}
\usepackage{amssymb}
\usepackage{amsbsy}
\usepackage{graphicx}
\usepackage{paralist}

\def\kpc{\,\rm kpc}

\def\kms{\,\rm km\,s^{-1}}
\def\rad{\,\rm rad}
\def\ActionUnits{\,\rm kpc^{2}Myr^{-1}}
\def\percent{\text{ per cent}}

\begin{document}
\title[Angles and actions in an axisymmetric potential]{Angle-action estimation in a general axisymmetric potential}
\author[J. Sanders]{Jason Sanders\thanks{E-mail: jason.sanders@physics.ox.ac.uk}\\
Rudolf Peierls Centre for Theoretical Physics, Keble Road, Oxford OX1 3NP, UK}

\pagerange{\pageref{firstpage}--\pageref{lastpage}} \pubyear{2012}
\label{firstpage}
\maketitle


\begin{abstract}
The usefulness of angle-action variables in galaxy dynamics is well known, but their use is limited due to the difficulty of their calculation in realistic galaxy potentials. Here we present a method for estimating angle-action variables in a realistic Milky Way axisymmetric potential by locally fitting a St\"ackel potential over the region an orbit probes. The quality of the method is assessed by comparison with other known methods for estimating angle-action variables of a range of disc and halo-type orbits. We conclude by projecting the Geneva-Copenhagen survey into angle-action space.  
\end{abstract}



\section{Introduction}
In the study of dynamical systems it is becoming increasingly important to be able to process and understand large multi-dimensional data sets efficiently. The stars in our own galaxy, the Milky Way, are being increasingly observed in full six-dimensional phase-space through the combination of astrometry and radial velocity measurements. Full 6D phase-space information is currently available for stars in the solar neighbourhood from the Geneva-Copenhagen and RAVE surveys \citep{Nordstrom2004,Zwitter2008} and this is to be greatly expanded on by the future space mission {\it Gaia} \citep{Perryman2001}. Beyond our Galaxy, the advent of integral-field spectroscopy has led to projects such as SAURON \citep[and subsequent papers]{Bacon2001}, which mapped the kinematics of a representative sample of 72 nearby elliptical and spiral galaxies, and subsequently ATLAS3D \citep[and subsequent papers]{Cappellari2011}, which combined SAURON observations with CO and HI observations to study the kinematics of a complete volume-limited sample of 260 local early-type galaxies. 
Observational data is often understood by performing large N-body simulations. Whilst such models are straightforward to produce, the configurations of the models are difficult to control and characterise. Schwarzschild modelling offers an improvement on this by describing the configuration of a model by a weighted set of orbits. However, this approach is not the most natural as each orbit is characterised by its initial phase-space coordinates. It is necessary that techniques are developed which can simplify both observational and simulation data without losing the richness of the phase-space information.

Angle-action variables are a set of canonical coordinates which can be used to express the equations of motion in a trivial form: the actions are integrals of the motion whilst the angles increase linearly with time. Such a formulation instantly reduces the complexity of any dynamical data set by reducing the six phase-space dimensions to three angle coordinates. Angle-action variables can be defined for any quasi-periodic orbit. Initially introduced to study celestial mechanics, angle-action variables now have great potential for galaxy dynamics due to their attractive properties. For instance, the Jeans theorem states that the arguments for the distribution function of a steady-state galaxy must be integrals of the motion, and it is particularly convenient to use as integrals the actions as
\begin{inparaenum}[(i)]
\item they are adiabatic invariants,
\item the zero-point of an action is well defined and
\item the range of values an action may take is independent of the other actions.
\end{inparaenum}
 The angle-action variables also provide a basis for the development of a perturbative solution to the equations of motion (see \citealt{BT08} for a much fuller discussion of the merits of angle-action variables).

The increasing evidence of substructure within the stellar halo of the Milky Way \citep[e.g.][]{Belokurov2006} has led many authors to consider the use of angle-action variables when hunting for and understanding the formation of structure within phase-mixed data sets. For example, \citet{McMillanBinney2008} studied a simulation of a self-gravitating satellite in a realistic Galaxy potential in angle-action space. Though the stars became well phase-mixed, the action space still showed considerable structure and through the use of angle-variable diagnostics the Galaxy potential and history of the satellite could be reconstructed. Similarly, \citet{Sellwood2010} and \citet{McMillan2011} have used the Geneva-Copenhagen survey to analyse stars in the solar neighbourhood in angle space and showed that the Hyades moving group may be due to a recent inner or outer Lindblad resonance. The study of tidal streams has a natural expression in angle-action variables. Under certain conditions the dimensionality of the stream may be reduced to one as the stream stretches out in a single angle coordinate \citep{Tremaine1999}. \citet{EyreBinney2011} showed that the path of a stream can be reconstructed far more reliably in angle-action space than by incorrectly assuming that streams delineate the orbits of their progenitors.

Despite the aforementioned advantages, angle-action variables remain awkward to work with in practical applications due to the difficulty of their calculation in a general potential. They are easily calculated when the potential is spherical and with more work can be analytically calculated when the potential is of St\"ackel form, but neither of these approaches is satisfactory when working with realistic galaxy potentials, because such potentials do not satisfy these conditions. The development of methods to estimate angle-action variables in a general potential is crucial if we are to benefit from the advantages of angle-action variables and the wealth of techniques which utilise them. 

In this paper we present a method for estimating angle-action variables in a general axisymmetric potential. The method proceeds by fitting a St\"ackel potential locally to the region of the potential a given orbit explores, thus enabling us to calculate analytically the actions and angles in this fitted St\"ackel potential. In Section~\ref{AAinStackelPot} we give a brief overview of the determination of angle-action variables in an axisymmetric St\"ackel potential and then in Section~\ref{StackFit} present the method for locally fitting such a potential to any axisymmetric potential. The results of the method are examined by analysing artificial data in Section~\ref{Application} and then these results are compared to other methods in Section~\ref{OtherMethodComparison}. Finally we demonstrate the practical application of the method by inspecting the Geneva-Copenhagen Survey in angle-action space in Section~\ref{GCS}.


\section{Actions and Angles in a St\"ackel potential}
\label{AAinStackelPot}
The most general class of potentials in which we are able to calculate the angle-action variables analytically is that of St\"ackel potentials. In a confocal ellipsoidal coordinate system these potentials produce separable Hamilton-Jacobi equations. A full discussion of St\"ackel potentials is given in \citet{deZeeuw1985a}. Here we limit the discussion to oblate axisymmetric St\"ackel potentials which are associated with prolate spheroidal coordinates $(\lambda,\phi,\nu)$. A specific prolate spheroidal coordinate system is defined by two constants $(a,c)$. These coordinates are related to cylindrical polar coordinates $(R,\phi,z)$ by
\begin{equation}
\frac{R^2}{\tau-a^2}+\frac{z^2}{\tau-c^2} = 1,
\end{equation}
where $\lambda$ and $\nu$ are the roots of $\tau$ such that $c^2\leq\nu\leq a^2\leq\lambda$. Surfaces of constant $\lambda$ are prolate spheroids and surfaces of constant $\nu$ are two-sheeted hyperboloids of revolution which intersect the spheroids orthogonally. A potential, $\Phi_S$, is of St\"ackel form in a particular prolate spheroidal coordinate system if 
\begin{equation}
\Phi_S = -\frac{f(\lambda)-f(\nu)}{\lambda-\nu}.
\label{StackDef}
\end{equation}
$\Phi_S$ is fully defined by a single function $f(\tau)$. A single function may be used as $\lambda$ and $\nu$ take different ranges of values except at $\lambda=a^2, \nu=a^2$, where we require $f$ to be continuous so the potential remains finite.
As in all axisymmetric potentials, the energy, $E$, and $z$-component of the angular momentum, $L_z$ are isolating integrals. In a St\"ackel potential we are in the fortunate position of being able to find  analytically a third isolating integral, $I_3$:
\begin{equation}
I_3 = (\lambda-c^2)\Bigg(E-\frac{L_z^2}{2(\lambda-a^2)}+\frac{f(\lambda)}{\lambda-c^2}-\frac{\dot{\lambda}^2(\lambda-\nu)^2}{8(\lambda-a^2)(\lambda-c^2)^2}\Bigg).
\label{I3}
\end{equation}
Therefore, given a Cartesian phase-space point $(\boldsymbol{x}, \boldsymbol{v})$, we can find the three isolating integrals, $\boldsymbol{I} = (E, L_z, I_3)$, using the coordinate transformation and calculating $\dot{\lambda}$ from $\boldsymbol{v}$.
Using the three isolating integrals we can write the first integrals of motion as
\begin{equation}
2(\tau-a^2)p_\tau^2 = E-\frac{L_z^2}{2(\tau-a^2)}-\frac{I_3}{\tau-c^2}+\frac{f(\tau)}{\tau-c^2},
\label{TauEqOfMotion}
\end{equation}
where $p_\tau$ is the momentum conjugate to $\tau=\lambda, \nu$.
We define the action variables, $J_\lambda$ and $J_\nu$, as
\begin{equation}
J_\tau = \frac{1}{2\pi}\oint p_\tau\mathrm{d}\tau,
\label{actionDef}
\end{equation}
where the integration is over all values of $\tau$ for which $p_\tau^2\geq0$. As $p_\tau = p_\tau(\tau, E, I_2, I_3)$ the actions are solely functions of the isolating integrals and thus constants of the motion. The third action, $J_\phi$, is simply $L_z$. The actions give an absolute measure of the extent of the oscillations of the orbit in each of the coordinates. At large radii the prolate spheroidal coordinate system becomes spherical such that $\lambda\approx R^2+z^2$. Therefore we can think of $J_\lambda$ as a measure of the radial oscillations. The $\nu$ coordinate increases as we move away from the $z=0$ plane so we may think of $J_\nu$ as a measure of the vertical oscillations.

The corresponding angle coordinates, $\theta_\tau$ and $\theta_\phi$, are calculated by the introduction of the generating function, $S(\lambda,\phi,\nu,J_\lambda,L_z,J_\nu)$ for the canonical transformation from $(\lambda,\phi,\nu,p_\lambda,p_\nu,L_z)$ to $(\theta_\lambda,\theta_\phi,\theta_\nu,J_\lambda,L_z,J_\nu)$. The angles are found by differentiating the generating function with respect to the respective action such that
\begin{equation}
\theta_\tau = \frac{\partial S}{\partial J_\tau} \mbox{	for	} \tau = \lambda,\nu;\> \theta_\phi=\frac{\partial S}{\partial L_z}.
\end{equation}
A full list of formulae, as well as a discussion of how to perform the quadratures numerically, is given in Appendix~\ref{AppendixAACalc}.


\section{Fitting a potential with a St\"ackel potential}
\label{StackFit}
Given the ease with which we can calculate actions and angles in a St\"ackel potential, it seems sensible to investigate how well a St\"ackel potential can fit a Galaxy model so that we may calculate the actions and angles in this best-fit potential. It has been known for some time that St\"ackel potentials do not give a good fit to the potential of the Galaxy globally due to the rigid conditions they must fulfil. \cite{DejonghedeZeeuw1988} outline a method for fitting an axisymmetric potential with a St\"ackel potential, which can be applied both globally and locally. These authors produced global fits for the Bahcall-Schmidt-Soneira Galaxy model (\citealt{BSS1982}) with errors nowhere exceeding $3\percent$, and \cite{Jasevicius1994} carried out a similar analysis on a broader range of Milky Way potential models with similar results. As expected, the fits are worst in the central $0.5\kpc$ of the Galaxy. \cite{DeBruyne2000} sought to fit axisymmetric potentials locally using a set of St\"ackel potentials in order to calculate the third isolating integral, $I_3$. When applied to a Miyamoto-Nagai potential, $I_3$ was found to vary by approximately $10\percent$ along an orbit. Here we follow the method presented by \cite{DejonghedeZeeuw1988}.

Suppose we have an axisymmetric potential $\Phi(R,z)$ that we wish to fit by a St\"ackel potential $\Phi_{\rm fit}$. We begin by choosing a prolate spheroidal coordinate system by specifying $(a,c)$. The coordinate system is fully specified by the combination $(a^2-c^2)$ so we are free to set $c^2 = 1$, which reduces numerical difficulties. We determine $a$ by using a property of an axisymmetric St\"ackel potential (\citealt{deZeeuwThesis}). 
It follows from equation \eqref{StackDef} that for a St\"ackel potential, $\Phi_S$,
\begin{equation}
\frac{\partial^2}{\partial\lambda\partial\nu}[(\lambda-\nu)\Phi_S]=0.
\end{equation}
Therefore, for a general potential $\Phi$ we use this equation as a definition for the coordinate system. Using expressions for $R$ and $z$ as a function of $(\lambda,\nu)$,
\begin{equation}
\begin{split}
R^2  &= \frac{(\lambda-a^2)(\nu-a^2)}{c^2-a^2},\\
z^2  &= \frac{(\lambda-c^2)(\nu-c^2)}{a^2-c^2},
\end{split}
\end{equation}
we find this gives an estimate for $a$ at a point $(R,z)$
\begin{equation}
a^2-c^2 = R^2-z^2-\Big[3z\frac{\partial \Phi}{\partial R}-3R\frac{\partial \Phi}{\partial z}+Rz\Big(\frac{\partial^2 \Phi}{\partial R^2}-\frac{\partial^2 \Phi}{\partial z^2}\Big)\Big]/\frac{\partial^2 \Phi}{\partial R\partial z}.
\label{DeltaGuess}
\end{equation}
We calculate a sufficiently accurate value of $a$ by evaluating this expression at multiple positions along the orbit and averaging.
With this choice of $a$ we transform $\Phi(R,z)$ to $\Phi(\lambda,\nu)$ and specify the fitting region: $\lambda_-\leq\lambda\leq\lambda_+$, $\nu_-\leq\nu\leq\nu_+$. A global fit corresponds to $\nu_-=c^2, \nu_+=\lambda_-=a^2,\lambda_+=\infty$. We also define the {\it auxiliary function} 
\begin{equation}
\chi(\lambda,\nu) \equiv -(\lambda-\nu)\Phi(\lambda,\nu).
\end{equation} 
If the potential $\Phi$ is of St\"ackel form, this auxiliary function is simply $\chi(\lambda,\nu) = f(\lambda)-f(\nu)$. We seek the function $f$ which makes $\Phi_{\rm fit}$ most like $\Phi$ by minimising the square difference of the potential auxiliary function and the fitting potential auxiliary function, $\chi_{\rm fit}$, over the fit region. Therefore, we minimise the functional
\begin{equation}
F[f] = \int_{\lambda_-}^{\lambda_+}\mathrm{d}\lambda\int_{\nu_-}^{\nu_+}\mathrm{d}\nu\,\Lambda(\lambda)N(\nu)(\chi(\lambda,\nu)-f(\lambda)+f(\nu))^2,
\label{leastsq}
\end{equation}
where $\Lambda(\lambda)$ and $N(\nu)$ are weighting functions allowing us to acquire a better fit in certain areas. These functions must be 
finite when integrated over the fitting region. 
We choose the normalised weighting functions 
\begin{equation}
\Lambda(\lambda)=4\lambda^{-5}(\lambda_-^{-4}-\lambda_+^{-4})^{-1},\>\>\> N(\nu) = (\nu_+-\nu_-)^{-1}.
\end{equation}
This choice of weighting functions gives preferential weight to smaller values of $\lambda$ where the potential is harder to fit.
Analytic minimisation of the functional $F$ results in a best fit function
\begin{equation}
f(\lambda) = \bar{\chi}(\lambda) - \frac{1}{2}\bar{\bar{\chi}},\>\> f(\nu) = -\bar{\chi}(\nu) + \frac{1}{2}\bar{\bar{\chi}}
\label{BestFitF}
\end{equation}
where
\begin{equation}
\begin{split}
\bar{\chi}(\lambda) &= \int_{\nu_-}^{\nu_+}\mathrm{d}\nu\,\chi(\lambda,\nu)N(\nu)\\
\bar{\chi}(\nu) &= \int_{\lambda_-}^{\lambda_+}\mathrm{d}\lambda\,\chi(\lambda,\nu)\Lambda(\lambda)\\
\bar{\bar{\chi}} &= \int_{\nu_-}^{\nu_+}\int_{\lambda_-}^{\lambda_+}\mathrm{d}\lambda\mathrm{d}\nu\,\chi(\lambda,\nu)\Lambda(\lambda)N(\nu).
\end{split}
\label{chibarFormulae}
\end{equation}
The derivation of these equations is given in Appendix~\ref{BestFitStackAppendix}. The quality of the fit we have achieved is then measured by $F[f]$. 
 

\section{Procedure}
Combining the above two sections we can estimate the actions and angles of a phase point $(\boldsymbol{x}, \boldsymbol{v})$ by first fitting a St\"ackel potential to the given potential over the region the orbit probes and then calculating the angle-action variables in this fitted potential. Therefore, given a point $(\boldsymbol{x}, \boldsymbol{v})$ we follow this procedure:
\begin{enumerate}
\item We begin by calculating the $z$-component of the angular momentum, $L_z$, and the energy, $E$, in the `true' potential.
\item We then integrate the orbit in the `true' potential. We use the initial time-steps of the orbit integration to find the best-fit coordinate system: at several points along the orbit we evaluate equation \eqref{DeltaGuess} and average to find a sufficiently accurate value for $a$.
With the coordinate system found, we continue integrating to find the edges of the orbit $\lambda_+, \lambda_-$ and $\nu_+$, which define the fitting region. The edges of the orbit are given approximately by the points where $\dot{\tau}=0$ in the best-fit ellipsoidal coordinate system.
The minimum and maximum $\tau$ edges are distinguished by inspecting the sign of $\ddot{\tau}$. 
For all realistic potentials every orbit crosses the $z=0$ plane so we set $\nu_-=c^2$.
\item We can now find a best-fit St\"ackel potential over this region. Using equation \eqref{BestFitF} we tabulate $f(\lambda)$ and $f(\nu)$ for 40 points in $(\lambda_-,\lambda_+)$ and $(\nu_-,\nu_+)$ respectively so that we may interpolate these smooth functions. Any call outside the ranges is calculated fully using equation \eqref{BestFitF} with a full re-computation of $\bar{\chi}(\tau)$.
\item With the best fit potential now calculated, we find $I_3$ using equation \eqref{I3} for three points on the boundary of the orbit (on the minimum $\lambda$ edge, the maximum $\lambda$ edge and the maximum $\nu$ edge) and take an average. We have already found these three points when determining the edges of the orbit so this choice involves minimum additional computational effort and provides a fair estimate for $I_3$ over a large region of the orbit. However, this choice of $I_3$ can lead to the initial phase-space point $(\boldsymbol{x}, \boldsymbol{v})$ being forbidden. Therefore, with this choice of $I_3$ we check whether $p^2(\nu)>0$ and $p^2(\lambda)>0$ for the initial phase-space point using equation~\eqref{TauEqOfMotion}, and if not then we calculate $I_3$ from equation~\eqref{I3} using only the initial phase-space point. This procedure reduces the numerical noise around the turning points, particularly in $R$.

\item With the three isolating integrals calculated, we are in a position to estimate the actions and angles using the method outlined in Section~\ref{AAinStackelPot}. The limits of the orbit are redetermined by finding from equation \eqref{TauEqOfMotion} the points where $p_\tau^2=0$ and are not given by $\tau_\pm$.
\end{enumerate}
Table~\ref{ErrorTimeData} quantifies the efficiency of this procedure.

\section{Application}
\label{Application}
\begin{figure}
$$\includegraphics[scale=1.0]{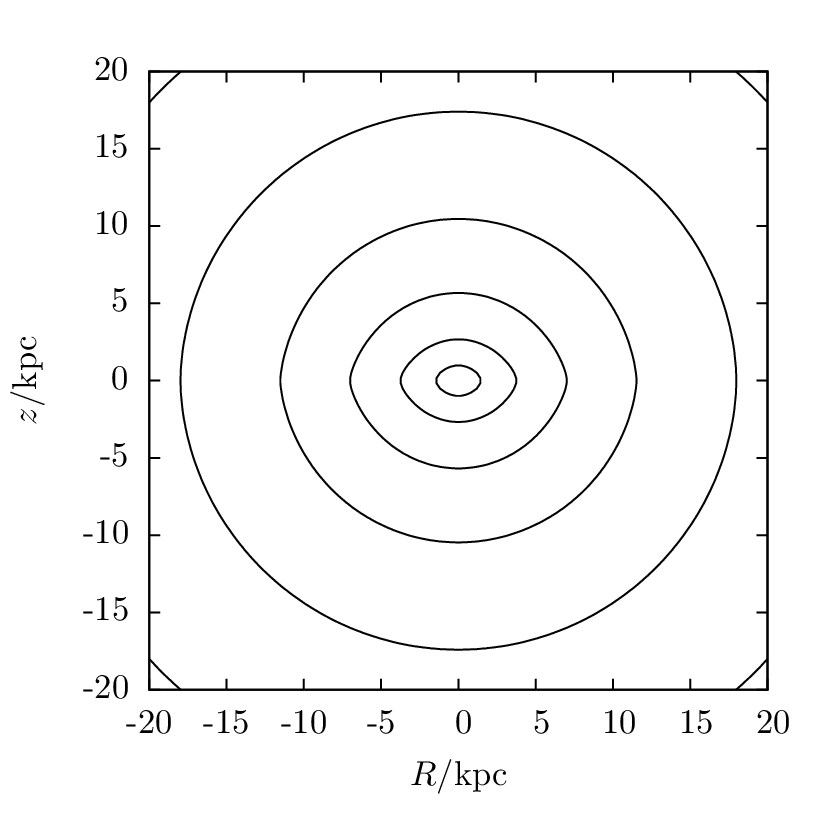}$$
\caption{Contours of $\ln(\Phi(R,z)/\Phi(0,0))$ for McMillan's best-fit Milky Way potential. The contours are increasing from the centre in equally space units of $0.15$ with the central contour at $-0.15$. }
\label{EquiPot}
\end{figure}

We now investigate how successful the above routine is in calculating the action-angle variables in a general axisymmetric potential. To demonstrate the applicability of the method to data we choose a realistic Milky Way potential from \citet{McMillan2011}. This potential consists of two exponential discs for the thick and thin discs of the Galaxy and two spheroids for the bulge and dark matter halo. 
We select the `best' model from this paper. The equipotential contours for this model are plotted in Figure~\ref{EquiPot}. It is clear that as we move out from the centre, the contours become more circular so we anticipate that they are better fit by surfaces of constant $\lambda$ and $\nu$. Therefore, we expect more accurate estimates of the angle-action variables for orbits at larger radii. Also orbits that probe a large range of $R$ and/or $z$ should have less accurate angle-action variable estimates as these orbits probe a large range of curvature of the equipotential contours. 
Therefore, we expect the method to work best for small $J_\lambda$ and $J_\nu$ but large $L_z$.

We assess the validity of the method by comparing the results with the `exact' angle-action variables calculated using the `torus machine' (\citealt{McMillanBinney2008}). Orbital tori are three-dimensional surfaces characterised by the three actions $\boldsymbol{J} = (J_R, L_z, J_z)$ obtained as the images of analytic tori under a canonical transformation. The strength of the torus machine lies in constructing a torus given a set of actions, $\boldsymbol{J}$, such that the phase-space coordinates, $(\boldsymbol{x}, \boldsymbol{v})$, may be obtained as functions of the angles, ${\boldsymbol{ \theta}}$, over the surface of the torus. Therefore, a simple test for the St\"ackel potential fitting procedure is to produce a list of phase space coordinates with fixed actions but randomly chosen angles using the torus machine. The success of the method is then measured by how accurately the angle-action variables can be reproduced. We note that the canonical transformation produced by the torus machine maps $J_R$ into $J_\lambda$ and $J_z$ into $J_\nu$. From now on we will use the more intuitive notation for the actions, $J_R$ and $J_z$, and similarly for the angles, $\theta_R$ and $\theta_z$.

The errors in the actions of a given torus from the torus machine may be estimated from the residuals of the Hamiltonian over the surface of the torus. The error in the Hamiltonian, $\Delta H$, is related directly to the error in one of the actions by 
\begin{equation}
\Delta H = \frac{\partial H}{\partial J}\Delta J = \Omega\,\Delta J
\end{equation}
where we find the frequency $\Omega$ directly from the torus machine. Assuming the errors in $J_R$ and $J_z$ are approximately equal and uncorrelated, the error in the actions may be estimated as
\begin{equation}
\Delta J\approx \frac{\Delta H}{\sqrt{\Omega_R^2+\Omega_z^2}}.
\end{equation}
The true angles of an orbit in a potential increase linearly with time. The errors in the torus angles are estimated by the residuals of the angles away from this expected straight line. Clearly we require these errors to be smaller than the errors from the St\"ackel fitting procedure in order to state anything meaningful about the systematic errors from our method.

\begin{figure}
$$\includegraphics[trim=0 0 10pt 20pt]{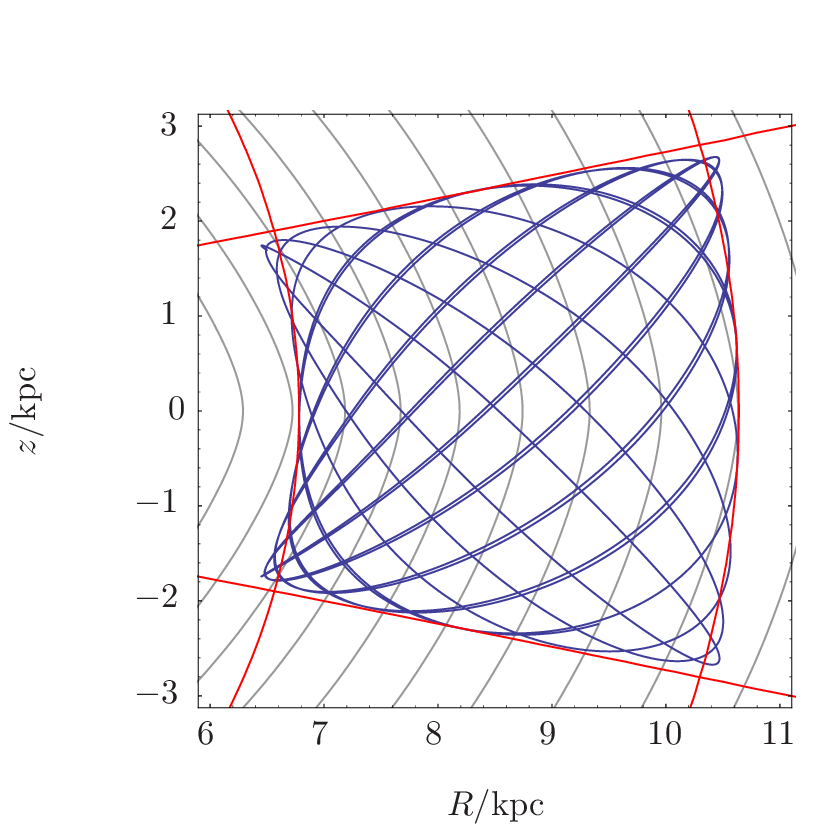}$$
\caption{Fit region for a single orbit - the blue line shows the orbit with actions $\boldsymbol{J} = (J_R, L_z, J_z) = (0.078, 1.9, 0.097)\ActionUnits$. The black lines are the equipotential contours of McMillan's best-fit Milky Way potential. The red lines show the lines of constant $\lambda$ and $\nu$ which define the region over which the potential is fitted. 
}
\label{FitRegionwOrbit}
\end{figure}

\subsection{Single Torus}
\label{SingleOrbitSection}
Here we discuss the results of applying the procedure to 10000 randomly generated points from the torus\footnote{Throughout this paper the actions are stated in units of $\ActionUnits = 977.8\kpc\kms$} $\boldsymbol{J} = (J_R, L_z, J_z) = (0.078, 1.9, 0.097)\ActionUnits$. This torus was chosen to be representative of the actions of a disc star in the solar neighbourhood. For this torus the errors in the actions and angles are $\Delta J/J = 0.01\percent$ and $(\Delta \theta_R, \Delta \theta_\phi,\Delta \theta_z) = (1.0,0.2,1.0)\times10^{-5}\rad$. The orbit in the $(R,z)$ plane is shown in Fig.~\ref{FitRegionwOrbit}. This orbit has apses at $R\approx(6.5,10.5)\kpc$ and $z_{\rm max}\approx2.8\kpc$. Also shown in the figure are the curves defining the fit region and equipotential contours for McMillan's best-fit potential. 

\begin{figure}
$$\includegraphics{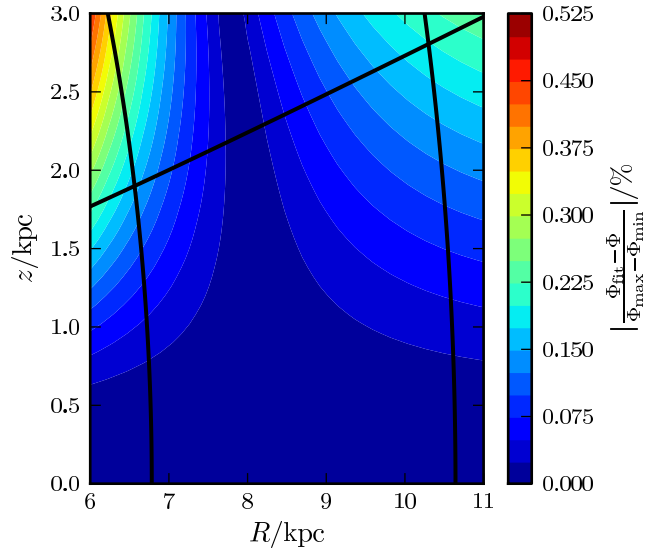}$$
\caption{Filled contour plot of the percentage difference between the best-fit St\"ackel potential and McMillan's best-fit Milky Way potential. $\Phi_{\rm min}$ and $\Phi_{\rm max}$ give the values of the potential on the minimum and maximum $\lambda$ edges respectively. Also plotted in black are the curves of constant $\lambda$ and $\nu$ which define the region over which the potential is fitted. This is the fit region corresponding to the orbit shown in Fig.~\ref{FitRegionwOrbit}, with actions $\boldsymbol{J} = (J_R, L_z, J_z) = (0.078, 1.9, 0.097)\ActionUnits$. We see that, within the fitting region, the difference between the fitted potential and the potential we are attempting to fit is less than $0.2\percent$ of the maximum potential difference across the fitting region.
}
\label{PotDiffFitRegion}
\end{figure}

The residuals in the fitted potential over the fitting region defined in Fig.~\ref{FitRegionwOrbit} are plotted in Fig.~\ref{PotDiffFitRegion}. Everywhere within the fitting region the error in the potential is less than $0.2\percent$ of the maximum difference in the potential across the fitting region. We note here that a good fit for the potential does not necessarily correlate with an accurate calculation of the actions. 
Small changes in the potential can cause large changes in the motion of a particle so, whilst a good fit for the potential is necessary, we don't expect the errors in the actions to be of similar order.

\begin{figure}
$$\includegraphics{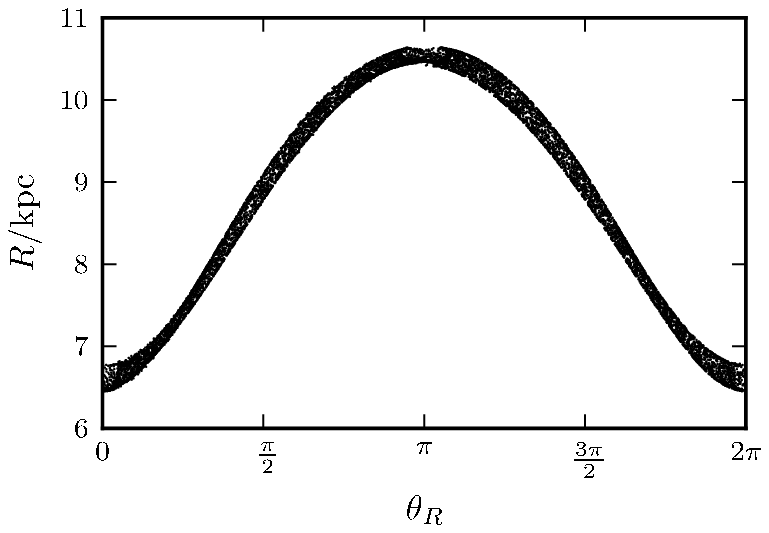}
$$\includegraphics{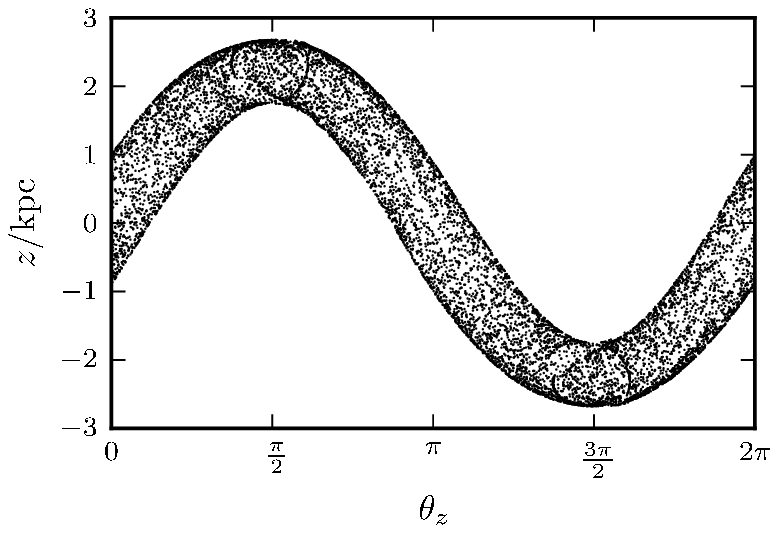}
\caption{Scatter plot of $R$ against $\theta_R$ and $z$ against $\theta_z$ for 10000 randomly selected phase-space points from the torus detailed in Section~\ref{SingleOrbitSection}.}
\label{PositionsAgainstAngles}
\end{figure}

The 10000 phase-space points are shown in scatter plots of $(R,\theta_R)$ and $(z,\theta_z)$ in Fig.~\ref{PositionsAgainstAngles}. We can see that $R$ and $z$ are periodic in the angles. We define the zero-point of $\theta_R$ such that the radial periapsis and apoapsis correspond to $\theta_R=0$ and $\theta_R=\pi$ respectively. $\theta_z$ is defined such that $z=0$ corresponds to $\theta_z=0,\pi$ and $z=\pm z_{\rm max}$ corresponds to $\theta_z=\pi/2,3\pi/2$. The zero-point of $\theta_\phi$ is defined such that $\theta_\phi=\phi$ at periapsis. The spread of the $z$ coordinates of the points at a given angle is much larger than the spread in the $R$ coordinates. 

When the St\"ackel fitting method is applied to this set of phase-space points, we find that the root-mean-square (RMS) deviations of the actions, $\Delta J$, are given by $\Delta J_R/J_R \approx 4.9\percent$ and $\Delta J_z/J_z \approx 4.2\percent$. We also find that there is a very tight anticorrelation between $J_R$ and $J_z$. All phase-space points have the same energy as we are using the potential that was used to integrate the orbit to find the energy. Therefore, all the points lie along the intersection of the surface of constant energy with the $(J_R,J_z)$ plane. If we overestimate $J_R$ then we must underestimate $J_z$ in order to have the correct energy. The RMS deviations in the angles for the 10000 phase-space points are $(\Delta \theta_R, \Delta \theta_\phi,\Delta \theta_z) = (4.1,1.1,5.1)\times10^{-2}\rad$.


\begin{figure}
$$\includegraphics{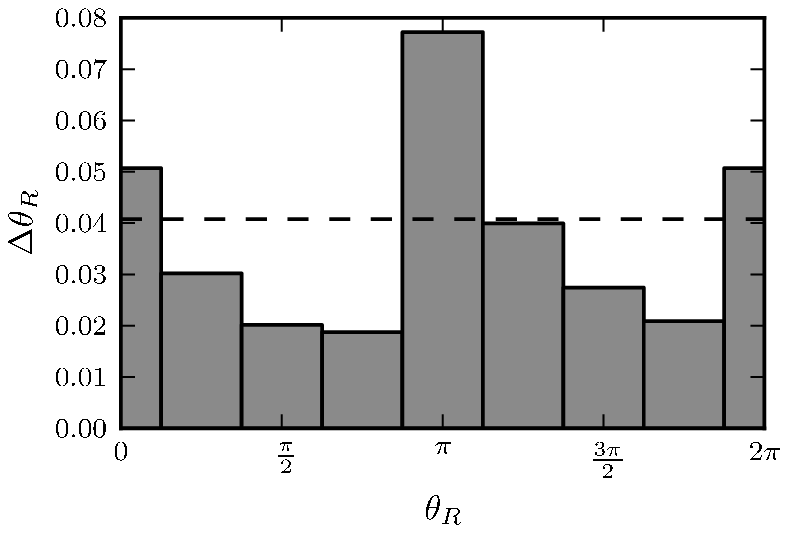}$$
$$\includegraphics{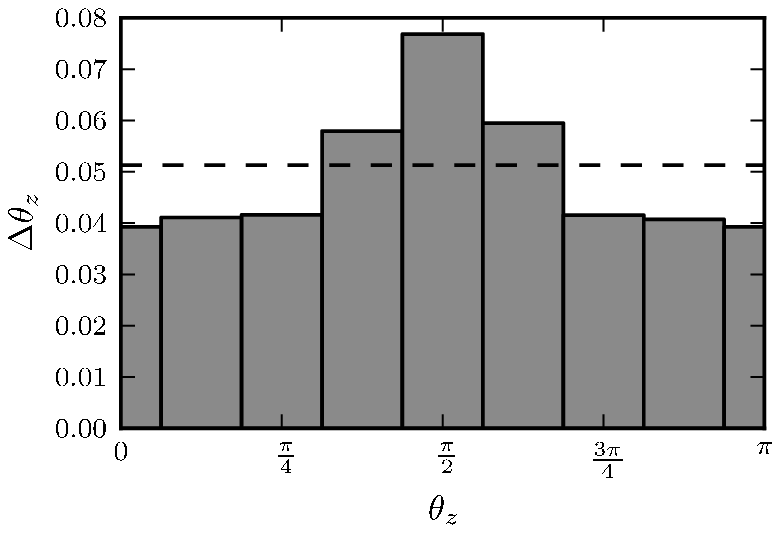}$$
\caption{RMS error in the angles binned as a function of angle. The dashed line shows the total RMS error from all the points on the torus. For the bottom panel we have taken advantage of the symmetry in the $z=0$ plane and mapped the $\theta_z=(0,2\pi)$ interval onto $\theta_z=(0,\pi)$ such that $\theta_z=\pi/2$ corresponds to $\pm z_{\rm max}$ etc. The largest error occurs at the apses for both cases.}
\label{ErrorswPosition}
\end{figure}

\subsubsection{Errors as a function of angle} 
It is informative to investigate how the errors in the derived actions and angles vary with true angle around the torus. The derived actions are approximately independent of the true angles as they depend only on the path of the orbit, which is determined by the fitted potential and not the initial point on the orbit. Any small variation is due to the choice of ellipsoidal coordinate system and variations in the fitted potential. However, we find that the error in the derived angle varies with true angle. 
In Fig.~\ref{ErrorswPosition} we plot the RMS errors in the angles binned as a function of true angle for both $\theta_R$ and $\theta_z$.
 Maximum errors occur at the turning points in the $(R,\theta_R)$ and $(z,\theta_z)$ plots shown in Fig.~\ref{PositionsAgainstAngles}. For the radial and vertical angle the largest error occurs at apoapsis.
In a St\"ackel potential the momenta, $p_\tau$, depend on $\tau$ and the isolating integrals, which once determined are taken to be constant. Therefore at a given location in the orbit the angle is solely a function of the position coordinates and the velocity information is essentially ignored. Around turning points in the orbit the velocity coordinates contain the majority of the information whilst the position coordinates are changing very slowly. Therefore at turning points the errors in the angles are large as the angle coordinates are estimated using this reduced phase-space information.
In general the errors in $\theta_z$ are larger than the errors in $\theta_R$. 

\subsection{Multiple Tori}
\label{MultipleOrbits}
We have seen that the method gives reasonable estimates for the actions for a particular torus, but in order to use the method with confidence we need to see how the errors depend on the torus. Here we repeat the above procedure for a range of different tori which probe the different regions of the potential. We work with two groups of tori: those with low actions and torus machine errors less than $\Delta J/J=0.01\percent$ and $(\Delta \theta_R, \Delta \theta_\phi,\Delta \theta_z) = (27.0,5.1,990)\times10^{-6}\rad$ and those with high actions and torus machine errors less than $\Delta J/J=1\percent$ and $(\Delta \theta_R, \Delta \theta_\phi,\Delta \theta_z) = (2.0,1.7,1.7)\times10^{-2}\rad$. The low-action group consists of 100 tori with actions $J_R = (0.001,0.005,0.01,0.05,0.1)\ActionUnits$, $J_z = (0.001,0.005,0.01,0.05,0.1)\ActionUnits$ and $L_z = (1.0,2.0,3.0,4.0)\ActionUnits$. These tori probe the region $3\kpc<R<22\kpc,\, z<5\kpc$ and are chosen to be representative of disc-type tori. The high-action group consists of 36 tori with actions $J_R = (0.5,1.0,5.0)\ActionUnits$, $J_z = (0.5,1.0,5.0)\ActionUnits$ and $L_z = (1.0,2.0,3.0,4.0)\ActionUnits$. These tori probe the region $2\kpc<R<120\kpc, \, z<100\kpc$. We include the second group to demonstrate that the method can deal with orbits which deviate very far from the plane and probe a very large region of the potential. We would like to be able to apply the method to halo stars and tidal streams so it is important to understand the errors for these high-action tori.

\subsubsection{Actions}
\begin{figure}
$$\includegraphics{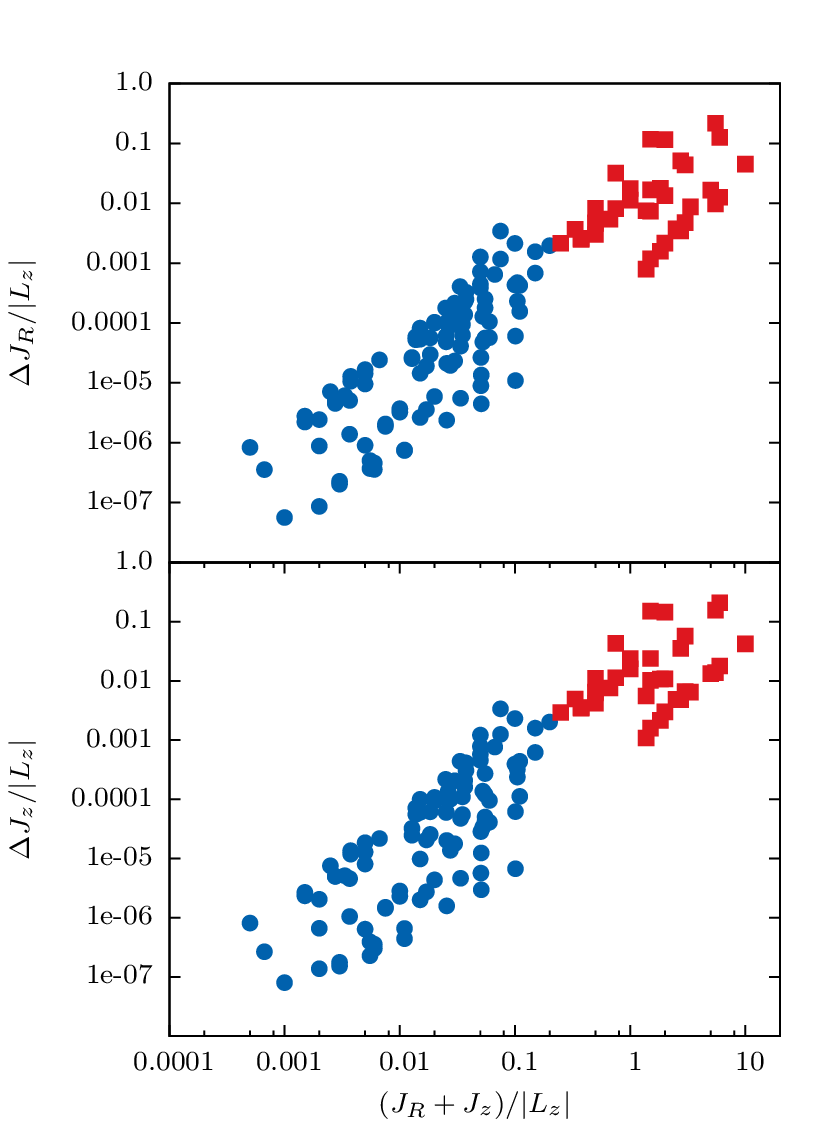}$$
\caption{Absolute RMS deviations of the actions. The blue circles are data for which the relative errors in the actions from the torus machine are less than $0.01\percent$ and the red squares are those with errors less than $1\percent$. The errors correlate loosely with both $J_R$ and $J_z$ separately, but there is a much tighter correlation between the errors and $(J_R+J_z)$.}
\label{JRLzJzJzLzAbsoluteErrors}
\end{figure}

As mentioned previously, we expect the errors in $J_R$ and $J_z$ will be large when $J_R/|L_z|$ and/or $J_z/|L_z|$ are large. In this regime the orbit probes a large central region of the potential so we anticipate the potential fit will be poorer. In Fig.~\ref{JRLzJzJzLzAbsoluteErrors} the RMS deviations in the actions for the complete orbit sample are plotted against the combination of the actions $(J_R+J_z)/|L_z|$. We can see that, as anticipated, the absolute errors correlate with this action combination. In fact, the correlation is much tighter than the individual correlations with $J_R/|L_z|$ and $J_z/|L_z|$, so the errors in the method are dependent on the sum of the actions $(J_R+J_z)$. It is this measure which tells us how much an orbit strays from a circular orbit and thus how much of the potential it explores. 

\begin{figure}
$$\includegraphics{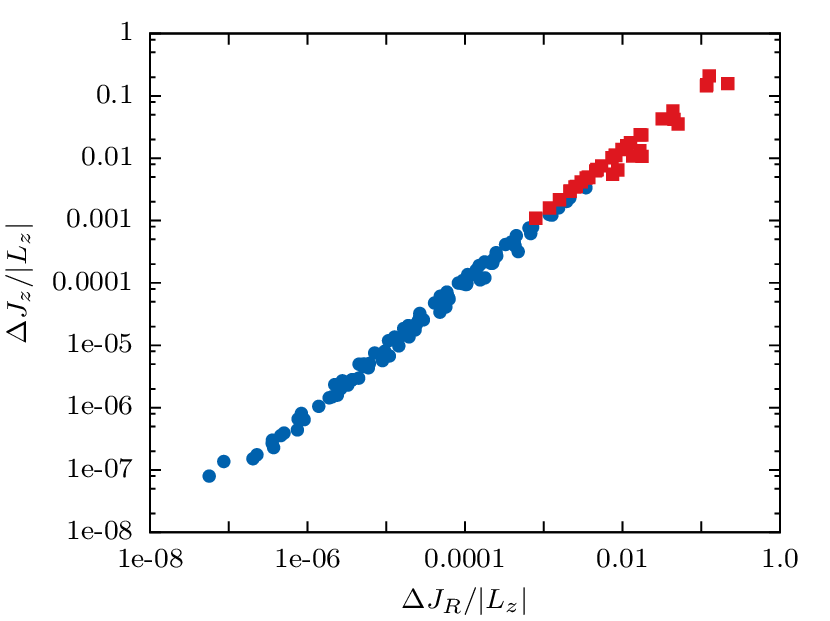}$$
\caption{RMS deviations in $J_R$ and $J_z$.  The blue circles are data for which the relative errors in the actions from the torus machine are less than $0.01\percent$ and the red squares are those with errors less than $1\percent$.}
\label{JRJzErrorCorrelation}
\end{figure}

We also note from Fig.~\ref{JRLzJzJzLzAbsoluteErrors} that at a given value of $(J_R+J_z)/|L_z|$ the errors in $J_R$ and $J_z$ are of similar magnitudes. As explained above, the errors in $J_R$ and $J_z$ compensate for each other to recover the correct energy.
In Fig~\ref{JRJzErrorCorrelation} we plot this correlation between the RMS errors in $J_R$ and $J_z$. A consequence of this tight correlation is that when one action is much greater than the other, the relative error in the smaller action will be much greater than the relative error in the larger action. However, it is worth noting that the absolute error is far more important than the relative error. Given a distribution function for a steady-state galaxy, $f(\boldsymbol{J})$, the absolute error in $f$ is given by 
\begin{equation}
(\Delta f)^2 = \sum_{i,j}\frac{\partial f}{\partial J_i}\frac{\partial f}{\partial J_j}{\rm cov}(J_i,J_j),
\end{equation}
where ${\rm cov}(X,Y)$ is the covariance between variables $X$ and $Y$. In the case of uncorrelated errors between the actions this simply becomes
\begin{equation}
(\Delta f)^2 = \sum_{i}\Big(\frac{\partial f}{\partial J_i}\Delta J_i\Big)^2.
\end{equation}
The distribution function for the Milky Way is approximately exponential in the actions \citep{Binney2010}:
\begin{equation}
f(\boldsymbol{J})\sim \prod_{i}\mathrm{e}^{a_iJ_i},
\end{equation}
where $a_i$ is independent of the action $J_i$. Therefore the absolute error in $f$ is given by
\begin{equation}
(\Delta f)^2 = \sum_{i}(f a_i \Delta J_i)^2.
\end{equation}
Similarly the relative error in the distribution function is given by
\begin{equation}
\Big(\frac{\Delta f}{f}\Big)^2 = \sum_{i}\Big(\frac{\partial \ln f}{\partial \ln J_i}\frac{\Delta J_i}{J_i}\Big)^2= \sum_{i}(a_i \Delta J_i)^2.
\end{equation}
Both the absolute and relative error in the distribution function are determined by the absolute errors in the actions, so we need not be overly concerned that the relative error in one action is much larger than the relative error in another.

From the relationship illustrated in Fig.~\ref{JRLzJzJzLzAbsoluteErrors} we can estimate the error in a given estimate of $J_R$ and $J_z$. Performing a linear fit to both sets of data points independently, we find that for $(J_R+J_z)/|L_z|\la 10$ a good fit for the RMS errors in both $J_R$ and $J_z$ is given by 
\begin{equation}
\Delta J\approx0.01\frac{(J_R+J_z)^{\frac{3}{2}}}{|L_z|^{\frac{1}{2}}}.
\end{equation}
The errors in the actions have a weak dependence on $L_z$ as orbits with higher $L_z$ explore regions of the potential which are more spherical and hence easier to fit with a St\"ackel potential (see Fig.~\ref{EquiPot}).

\subsubsection{Angles}
\begin{figure}
$$\includegraphics{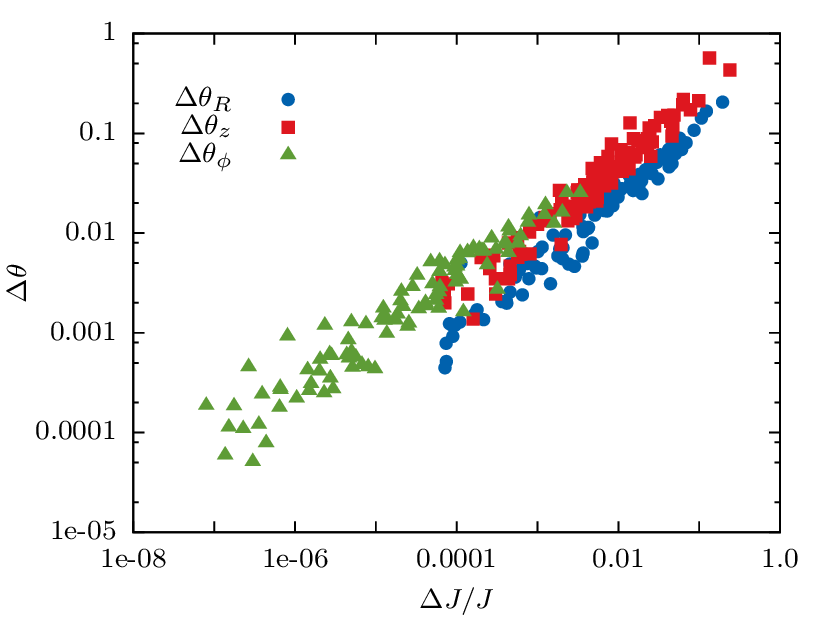}$$
\caption{RMS deviations in the angles for the 100 low-action tori as a function of the relative error in their respective action. For $\theta_\phi$ we have plotted the error against $\Delta J_R/L_z$.}
\label{AnglesAbsoluteErrors}
\end{figure}

We present the RMS deviations in the angles for the 100 tori in the low-action group in Fig.~\ref{AnglesAbsoluteErrors}.
In general the errors in the angles are larger than the errors in the actions. The calculation of an action involves only a single integral whereas the corresponding angle calculation involves nine integrals. Each integral folds in more error from the fitting method so we expect the errors in the angle variables to be significantly larger than the errors in the actions. 
From Fig.~\ref{AnglesAbsoluteErrors}  we see that the errors in the angles correlate with the relative error in the actions. The error in $\theta_\phi$ has been plotted against $\Delta J_R/L_z$ whilst the other two angles have been plotted against the relative error in their respective action. As the errors in $J_R$ and $J_z$ are approximately equal (Fig.~\ref{JRJzErrorCorrelation}) the three sets of points are essentially $\theta_i$ against $\Delta J/J_i$. As with the errors in the actions, we may estimate the error in a given calculation of the angles by fitting the data in Fig.~\ref{AnglesAbsoluteErrors}. We find the errors are approximately given by
\begin{equation}
\Delta\theta_{R}\approx\Big(\frac{\Delta J}{J_{R}}\Big)^{0.75},\>
\Delta\theta_{z}\approx\Big(\frac{\Delta J}{J_{z}}\Big)^{0.75},\>
\Delta\theta_{\phi}\approx\Big(\frac{\Delta J}{L_z}\Big)^{0.5}.
\end{equation}

\section{Comparison with other Methods}
\label{OtherMethodComparison}
\subsection{Total Angular Momentum}
\begin{figure}
$$\includegraphics{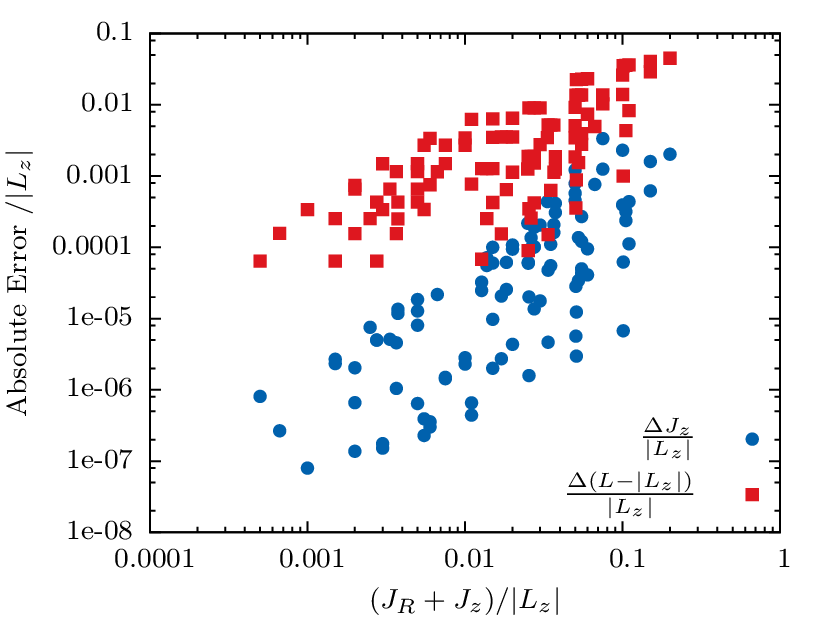}$$
\caption{RMS deviations of $J_z$ for the St\"ackel fitting method (blue circles) and the RMS deviations in the spherical vertical action $(L-|L_z|)$ (red squares) for the same sample of orbits. }
\label{TotalL}
\end{figure}

Some authors have hunted for structure in the distribution of stars in spaces defined by phase space functions other than the actions. For example, \citet{Helmi2000} use the set of variables $(E,L_z,L)$, where $L = |\boldsymbol{x}\times\boldsymbol{v}|$ is the total angular momentum, to attempt to find substructure within numerical simulations of disrupted satellite galaxies, whereas \citet{Helmi2006} considered the `APL space' of apocentre, pericentre and $z$-component of the angular momentum in order to identify signatures of past accretion events in the Geneva-Copenhagen Survey of the solar neighbourhood (\citealt{Nordstrom2004}). The total angular momentum is only conserved when we are considering spherical potentials. In a spherical potential the vertical action is simply $J_z=L-|L_z|$. Here we investigate how much better we are doing when we estimate the vertical action using a St\"ackel fit than if we simply use $L$. Fig.~\ref{TotalL} shows the absolute RMS error in the vertical action for the 100 low-action tori taken from the lower panel of Fig.~\ref{JRLzJzJzLzAbsoluteErrors} along with RMS error in the spherical vertical action, $(L-|L_z|)$. The St\"ackel fitting method gives approximately two orders of magnitude improvement in the vertical action error compared to simply using $L$.

\subsection{Adiabatic Approximation}
The adiabatic approximation provides an alternative method for estimating the actions. In its simplest form (\citealt{Binney2010}) this approximation assumes that the motion in the plane is unaffected by the motion perpendicular to the plane. The absence of energy transfer between the radial and vertical motion leads to an underestimate of the centrifugal potential for the radial motion and hence an underestimate of the maximum radius of an orbit. \citet{McMillanBinney2011} attempted to resolve this issue by replacing $L_z$ with $(|L_z|+J_z)$ in the effective radial potential. \citet{Schonrich2012} improved on this by including a correction to the radial energy due to the changes in the vertical energy along an orbit. It is this final approach that we test here.

Following \citet{Schonrich2012} we assume that the vertical motion at a given radius, $R_0$, is governed by the potential $\Psi_z(z) = \Phi(R_0, z)-\Phi(R_0,0)$ such that the vertical energy, $E_z$, is
\begin{equation}
E_z=\frac{1}{2}v_z^2+\Psi_z(z).
\end{equation}
Then the vertical action is estimated to be 
\begin{equation}
J_z = \frac{2}{\pi}\int_{0}^{z_{\rm max}}\mathrm{d}z\,v_z,
\end{equation}
where $z_{\rm max}$ is the point where the vertical velocity, $v_z$, is zero.
By linear interpolation we may reverse this calculation such that, for a given pair of $J_z$ and $R_0$, we may calculate $E_z(J_z,R_0)$. Over the course of an orbit we take $J_z$ to be constant but the vertical energy will be changing as the orbit explores different radii. For overall energy conservation this energy must be transferred from the vertical motion into the radial motion. Therefore, the radial motion is governed by the one-dimensional potential
\begin{equation}
\Psi_R(R) =  \Phi(R,0) + \frac{L_z^2}{2R^2}+E_z(J_z,R)-E_z(J_z,R_g),
\end{equation}
where $R_g$ is the guiding-centre radius (the radius of a circular orbit with $z$-component of angular momentum $L_z$).
Using this potential we estimate the radial action as
\begin{equation}
J_R = \frac{1}{\pi}\int_{R_p}^{R_a}\mathrm{d}R\,v_R,.
\end{equation}
where $R_p$ and $R_a$ are the points where the radial velocity, $v_R$, is zero.

\begin{figure}
$$\includegraphics{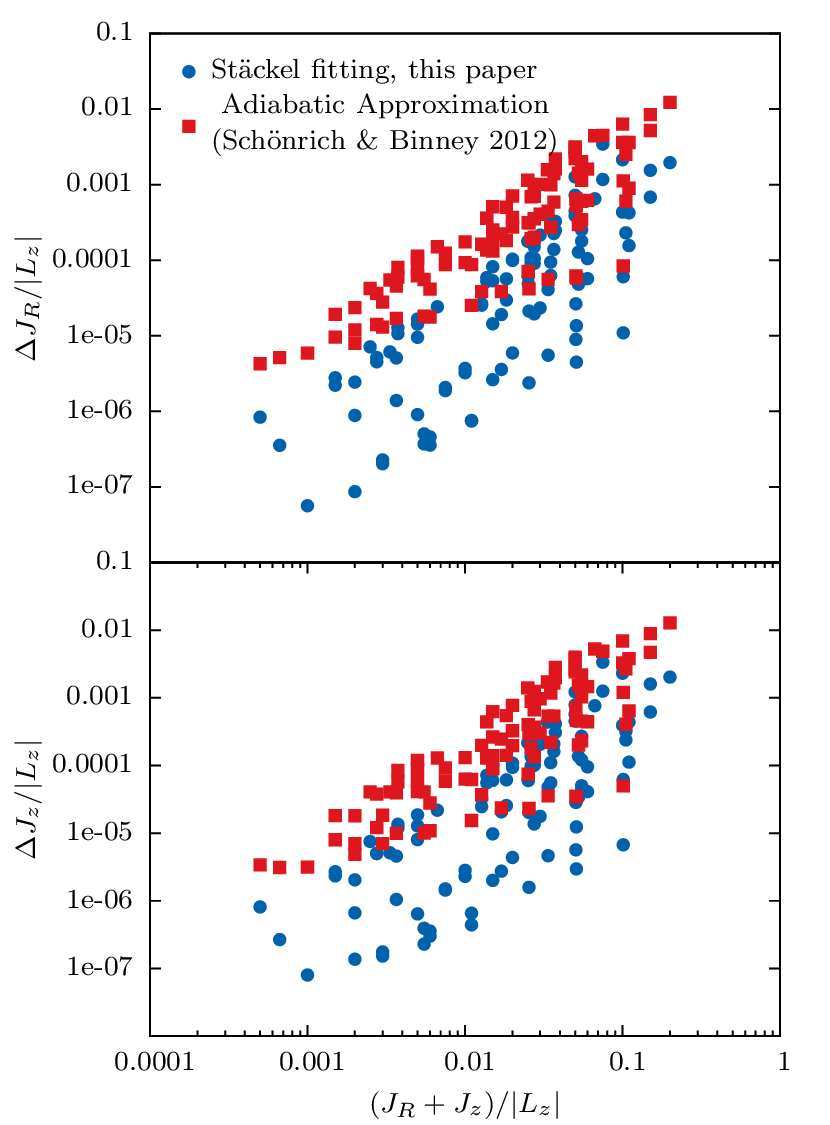}$$
\caption{Absolute RMS deviations of the actions for the St\"ackel fitting method (blue circles) and the adiabatic approximation (red squares). We can see that in general the St\"ackel fitting method is an improvement on the adiabatic approximation. At the low total action ($J_R+J_z$) end the improvement is around two orders of magnitude, whilst at the high total action end the improvement is reduced to less than one order of magnitude.}
\label{AbsErrorswithAAAA}
\end{figure}

This method is clearly much simpler than the full St\"ackel potential fitting method and for a single phase-space point is approximately two to twenty times faster. Here we will see whether the accuracy of the St\"ackel potential fitting method justifies its extra expenditure. In Fig.~\ref{AbsErrorswithAAAA} we have replotted the 100 low-action tori data from Fig.~\ref{JRLzJzJzLzAbsoluteErrors} along with the equivalent calculation using the adiabatic approximation. We find that for low total action, ($J_R+J_z$), the St\"ackel fitting method is about two orders of magnitude more accurate than the adiabatic approximation, whilst at the high total action end this improvement is reduced to less than one order of magnitude. It is also worth noting that the relative errors for the adiabatic approximation can be as much as order one for orbits with large $J_z$ - the adiabatic approximation performs best for orbits that stay near the plane. In Table~\ref{ErrorTimeData} we compare the errors in the actions and the time taken to estimate the actions for 10000 phase-space points for the St\"ackel fitting method and the adiabatic approximation. We see that for near-circular near-planar orbits we can achieve an order of two magnitude decrease in the errors of the actions using the St\"ackel fitting method whilst only incurring an additional cost of doubling the computational time. However, for orbits with much greater actions we can only achieve a single order of magnitude decrease in the error for a fivefold increase in the computational time.

\begin{table*}
\caption{Absolute RMS errors in the actions from the St\"ackel fitting method and the use of the adiabatic approximation along with the time taken to evaluate the actions of 10000 phase-space points for each method.}
\begin{tabular}{llllll}
$J_R/|L_z|$&$J_z/|L_z|$&$\Delta J^{\rm St\ddot ack}_{R}/\Delta J^{\rm AA}_{R}$&$\Delta J^{\rm St\ddot ack}_{z}/\Delta J^{\rm AA}_{z}$&Time$_{\rm St\ddot ack}$/s&Time$_{\rm AA}$/s\\
\hline
0.001 & 0.001 & 0.0087 & 0.12 & 10.1 & 4.5\\
0.01 & 0.01 & 0.023 & 0.028 & 15.1 & 6.4\\
0.1 & 0.1 & 0.16 & 0.16 & 44.6 & 8.3\\
\hline
\end{tabular}
\label{ErrorTimeData}
\end{table*}

\subsection{Iterative torus machine}
\label{IterativeTorus}
\citet{McMillanBinney2008} used the torus machine iteratively to find the actions and angles of a given phase-space point. This typically involved around 20 torus fits per phase-space point and relied on a good initial guess for the actions and angles for fast convergence. Such a method is potentially the most accurate way to determine the angle-action variables but also the most costly. The authors report that it takes around 15$s$ to perform the iterative procedure on a single phase-space point. The alternative method we present in this paper could complement this method as it provides an accurate initial guess for the actions and angles which may be fed into the torus machine. We are using the torus machine as a source of `true' actions and angles in order to test the accuracy of the methods, so we are assuming that, given enough iterations, this approach will give near-perfect results. Therefore, we do not explore the results of this method.

\section{Geneva-Copenhagen Survey}
\label{GCS}
\begin{figure}
$$\includegraphics{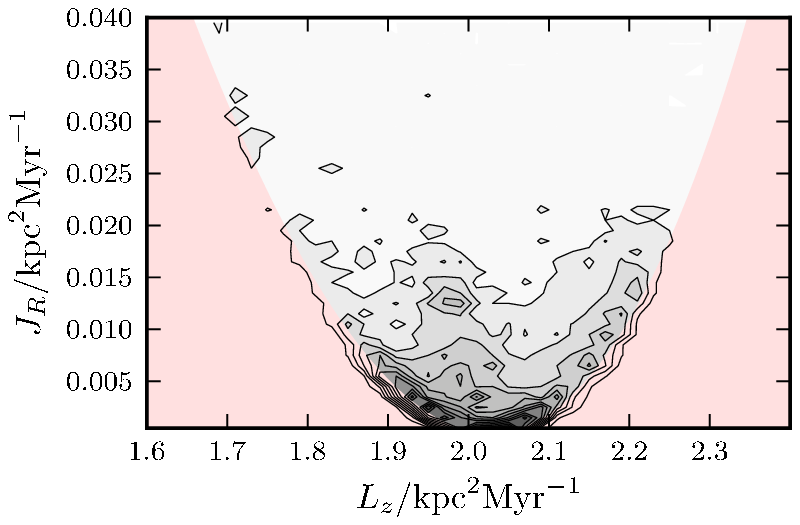}$$
$$\includegraphics{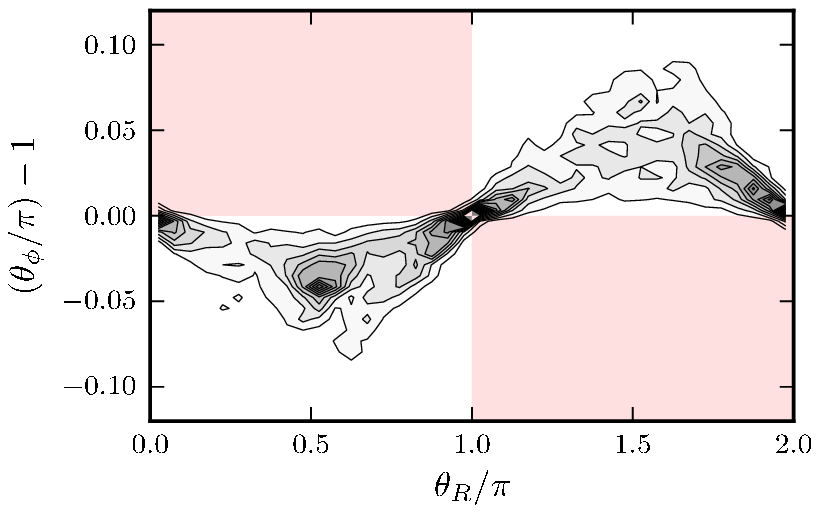}$$
\caption{Density contours for the actions and angles of 13518 objects from the Geneva-Copenhagen Survey. The contours are linearly spaced. The red shaded region is inaccessible by stars which are at the solar position and have $J_z=0$.}
\label{GCSResults}
\end{figure}

Now that we have understood the systematic errors of the method we apply it to real data. The Geneva-Copenhagen Survey (GCS) (\citealt{Nordstrom2004}) is a sample of 16682 nearby F and G stars, and is perhaps the best data set with full 6D phase-space information. It provides us with a platform to motivate a discussion of errors involved in a practical calculation. The GCS has been analysed by many authors and specifically looked at in angle-action space by \citet{Sellwood2010} and \citet{McMillan2011b}. From the table produced by \citet{Holmberg2009} we select the 13518 objects for which we have the full 6D phase-space information. We correct the data for the solar velocity with respect to the local standard of rest as calculated by \citet{Schonrich2010} i.e. $(U,V,W)_{\odot} = (11.1,12.24, 7.25)\kms$
\footnote{We work in a right-handed Galactocentric Cartesian coordinate system with the positive $x$ direction pointing towards the Galactic centre.}. 
Using the `best' model Galactic potential from \citet{McMillan2011} sets the solar radius as $R_\odot = 8.29\kpc$ and the velocity of the local standard of rest as $v_{\rm LSR} = 239.1\kms$. The results of estimating the actions and angles using the St\"ackel fitting procedure are shown in Fig.~\ref{GCSResults}. These are very similar to the equivalent plots from \citet{Sellwood2010} and \citet{McMillan2011b}. The shaded red region is inaccessible by stars which are at the solar position and have $J_z=0$.

We see that the plot of $J_R$ against $L_z$ has a markedly parabolic shape. The minimum of this parabola corresponds to the circular orbit at the solar radius. The edges of the parabola are defined by the limits of the survey. Stars with $L_z$ significantly different from the angular momentum of the local standard of rest, $L_{z,0} = R_\odot v_{\rm LSR}$, require a sufficiently large radial action to bring them into the solar neighbourhood. From the plot of $\theta_\phi$ against $\theta_R$ we see that the majority of stars are at $\theta_R=0,\pi$ corresponding to the apses of their radial motion. However, there is still a lot of structure in between these extrema: the peak at $\theta_R/\pi\approx0.54$ corresponds to the Hyades moving group. We do not plot the distribution in the vertical action and angle because these plots lack interesting features.

\subsection{Structure in the GCS}
\begin{figure}
$$\includegraphics{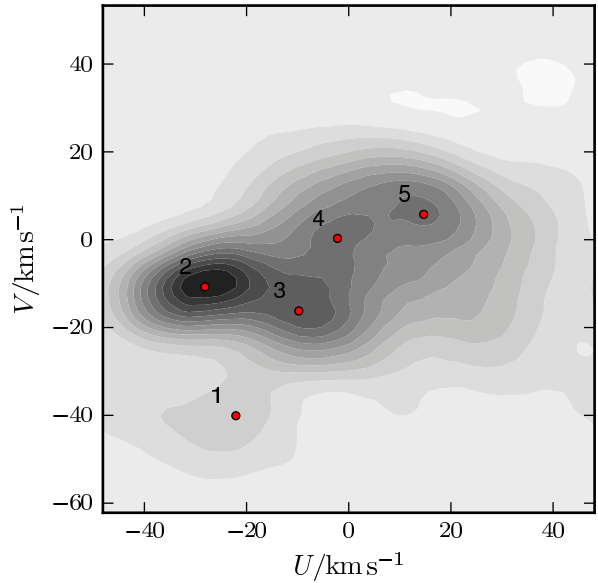}$$
\caption{Density contours for the $(U,V)$ plane of 13518 objects from the Geneva-Copenhagen Survey smoothed to a scale of $\sim 12\kms$ via a stationary wavelet transform. The five major structures are identified by red points: 1. Hercules, 2. Hyades, 3. Pleiades, 4. Coma Berenice, 5. Sirius.}
\label{UVContours}
\end{figure}

Since \citet{Dehnen1998} investigated the kinematics of the solar neighbourhood using data from the {\it Hipparcos} satellite, it has been known that when viewed in the $(U,V)$ velocity plane the local distribution of stars consists of a series of groups and clusters. These structures were classified by \citet{Famaey2005} and are all thought to have a dynamical origin \citep{DeSimone2004,Antoja2010}. From the GCS sample we can identify several of the larger structures from the peaks in the $(U,V)$ distribution. This is done by binning the stars in $U$ and $V$ and then performing a wavelet transform. In Fig.~\ref{UVContours} we show the results of the wavelet transform for structure on scales $\sim12\kms$. The peaks in this plot are identified with the known groups, specifically the Hercules, Hyades, Pleiades, Coma Berenice and Sirius. As discussed by \citet{McMillan2011b} for a star located at the solar position each point in the $(U,V)$ plane represents a unique point in the $(J_R,L_z,\theta_R,\theta_\phi)$ space. Lines of constant $L_z$ and $\theta_\phi$ form an approximately Cartesian grid in the $(U,V)$ plane, whilst lines of constant $J_R$ and $\theta_R$ form an approximately polar grid. Therefore, we can find a reduced set of angle-actions (excluding any vertical action and angle) for each of the peaks identified in Fig.~\ref{UVContours}. From inspecting the $(U,V)$ plane we can see that the distribution deviates from axisymmetric equilibrium, so estimating the actions and angles assuming an axisymmetric potential is clearly an oversimplification. However, this is a necessary first step to create a basis on which we can include non-axisymmetric perturbations.

We represent each group by a 6D phase-space point placed at the solar position with zero vertical velocity but $U$ and $V$ determined by the identification from Fig~\ref{UVContours}, and then estimate the actions and angles corresponding to this point, again using McMillan's `best' potential. The results are shown in Table~\ref{GroupCoordinates}. This gives us an opportunity to discuss the various sources of errors in a realistic use of the method. There are two sources of error in this calculation -- the systematic errors introduced by the St\"ackel fitting method and the errors in the input coordinates. \citet{Holmberg2009} estimated the space-velocity errors for each star as $1.5\kms$, which when combined with the errors estimated in \citet{Schonrich2010} for the solar motion gives velocity errors of $(\Delta U, \Delta V) = (1.7,1.6)\kms$. The majority of this error arises from the uncertainty in the distances. We convolve each peak position by these errors to calculate 10000 Gaussianly distributed points in $(U,V)$ space about these peaks, and then estimate the errors in the output actions and angles by the RMS scatter in the resulting $(\boldsymbol{J}, {\boldsymbol{\theta}})$ coordinates. These errors can then be compared and combined with the known systematic errors from Section~\ref{MultipleOrbits} and are shown in Table~\ref{GroupCoordinates}. 

We find that for all the coordinates, the error in the data dominates the systematic error introduced by the method. As noted by \citet{McMillan2011b} the relationship between errors in $(\boldsymbol{x}, \boldsymbol{v})$ and $(\boldsymbol{J}, {\boldsymbol{\theta}})$ is non-trivial. Specifically at very low radial actions, any error in the velocity can introduce a $2\pi$ error in the $\theta_R$ coordinate. We see this occurring for the Coma Berenice peak -- the error in the $\theta_R$ coordinate is large as the peak is positioned very close to the origin of the $(U,V)$ plane. We also see that the errors in $J_R$ and $\theta_\phi$ are of order one for Coma Berenice. 

We note that we have not included any error for the size of the structures in phase-space, nor any error for the assumption that all the stars are situated at the solar position, nor any error in the choice of gravitational potential. Investigating the error in the actions due to the range of viable potentials for the Milky Way is beyond the scope of this paper. Even with this underestimated error, the errors from the data dominate the systematic errors.
We conclude that, given the accuracy of the current data, the determination of the angle-action coordinates using the St\"ackel fitting method is not limited by the well-understood systematic errors. 

\begin{table*}
\caption{Velocities with respect to the local standard of rest and angle-action coordinates for density peaks of known structures in the Geneva-Copenhagen Survey. The errors in $U$ and $V$ are $1.7\kms$ and $1.6\kms$ respectively. The errors in the angles and actions are presented as $a_c^b$ where $a$ is the total error and $b$ and $c$ are the contributions from the systematic errors of the St\"ackel fitting method and the errors in the space-velocities respectively, such that $a^2=b^2+c^2$. In all cases the observational error dominates the systematic error.}
\begin{tabular}{lllllll}
\hline
&$U/\kms$&$V/\kms$&$J_R/\ActionUnits$&$L_z/\ActionUnits$&$\theta_R$&$\theta_\phi$\\
\hline
Hercules		&$-22.1$&$-40.1$&		$(4.0\pm0.3_{0.3}^{0.006})\times10^{-2}$	&		$1.69\pm{0.01}$&		$3.63\pm0.04_{0.036}^{0.008}$&		$(9.6\pm0.7_{0.68}^{0.05})\times 10^{-2}$\\

Hyades		&$-28.2$&$-10.8$&		$(1.2\pm0.1_{0.1}^{0.001})\times 10^{-2}$&		$1.94\pm{0.01}$&		$4.37\pm0.07_{0.07}^{0.005}$&		$0.118\pm0.007_{0.0066}^{0.0001}$\\

Pleiades		&$-9.8$&$-16.3$&		$(7.3\pm0.1_{0.1}^{0.005})\times 10^{-3}$&		$1.89\pm{0.01}$&		$3.61\pm0.08_{0.077}^{0.004}$&		$(4.2\pm0.7_{0.7}^{0.006})\times 10^{-2}$\\

Coma Berenice&$-2.2$&$0.3$&	$(1.7\pm1.4_{1.4}^{0.0002})\times 10^{-4}$&		$2.03\pm{0.01}$&		$5.0\pm1.1_{1.1}^{0.0009}$&			$(9.3\pm6.4_{6.4}^{0.0008})\times 10^{-3}$\\

Sirius		&$14.7$&$5.7$&		$(3.5\pm0.8_{0.8}^{0.001})\times 10^{-3}$&		$2.07\pm{0.01}$&		$1.01\pm0.13_{0.13}^{0.003}$&		$6.22\pm0.01_{0.006}^{0.00002}$\\
\hline
\end{tabular}
\label{GroupCoordinates}
\end{table*}

\subsection{Comparison with \citet{McMillan2011b}}
\citet{McMillan2011b} calculated the angles and actions of the GCS sample by using the torus machine iteratively. Here we compare the results to our own estimates of the angle-actions. The potential used by \citet{McMillan2011b} was the `convenient' potential detailed in \citet{McMillan2011}, which places the Sun at a galactocentric radius $R_\odot=8.5\kpc$ with a velocity of the local standard of rest $v_{\rm LSR} = 244.5\kms$. We calculate the RMS deviations between McMillan's data and ours, and present the results in Table~\ref{McMillanDiff}. We also show the expected RMS errors in the angles and actions estimated using the St\"ackel fitting method. The largest discrepancy between our data and McMillan's occurs for the $\theta_z$ coordinate. The expected error for this coordinate is also large, and very much larger than the actual error. This is because the sample is dominated by stars with low vertical action as the sample is situated in the Galactic plane. As the absolute error in the vertical action depends on the sum of the radial and vertical actions a large radial action can lead to a large relative error in the vertical action. The systematic error in $\theta_z$ depends on this large relative error. For all the other variables the discrepancy between our data and McMillan's data seems to be in agreement with the expected systematic errors of our method. 
\begin{table}
\caption{RMS deviations between actions and angles for stars in the Geneva-Copenhagen Survey sample estimated with the St\"ackel fitting method and the data from \citet{McMillan2011b}. The second column gives the RMS of the expected systematic errors from the St\"ackel fitting method.}
\begin{tabular}{r r r}
&\begin{tabular}[x]{@{}l@{}}RMS\\difference\end{tabular}
&\begin{tabular}[x]{@{}l@{}}Expected RMS\\error\end{tabular}\\
\hline
$\Delta J_R/10^{-4}\ActionUnits$	& 5.4		&4.9\\
$\Delta J_z/10^{-4}\ActionUnits$ 	& 7.6		&4.9\\ 
$\Delta \theta_R / 0.01\rad$		& 2.6		&0.9\\
$\Delta \theta_\phi / 0.01\rad$		& 1.2		&0.6\\
$\Delta \theta_z / 0.01\rad$		& 7.7		&40.1\\
\hline
\end{tabular}
\label{McMillanDiff}
\end{table}

\section{Conclusions}
We have detailed a method for estimating the angle-action variables in a general axisymmetric potential given a 6D phase-space point. The method is based on locally fitting a St\"ackel potential to the region of the potential the orbit probes and then taking advantage of the ease with which we may calculate the actions and angles in a St\"ackel potential. We have investigated the systematic errors by producing phase-space points of known actions and angles using the torus machine and then assessing how well the method can reproduce these variables. For a single torus the errors in the angles are largest for phase-space points near apsis and the errors in the actions are constant (of order a few percent). For a collection of tori, chosen to be representative of both disc and halo-type tori, the absolute error in the actions is found to scale with the sum of the vertical and radial actions. The errors in the angles scale with the relative error in their corresponding action. We compared the method to other methods for estimating the actions in an axisymmetric potential. The method gives results approximately two orders of magnitude more accurate than assuming the potential is spherical and performs approximately a single order of magnitude better than the adiabatic approximation. 

We have demonstrated that the procedure is suitable for most disc and halo-type orbits. The procedure will not work for resonant orbits or chaotic orbits. However, the occurrence of these orbits in realistic galaxy axisymmetric potentials is rare and the great majority of stars are on quasi-periodic non-resonant orbits \citep{Ollongren1962,MartinetMayer1975}.

We demonstrated the use of the method by application to the Geneva-Copenhagen Survey (GCS). As this survey is only local, the angle-action space does not reveal much more information than velocity space. However, we present angle-action coordinates for the peaks of the clumps and streams present in the survey and use them to study the relative impact on estimated angles and actions of observational errors and the known systematic errors of the method. We show that the observational errors are dominant.

It is hoped that this method will lead to more widespread use of angle-action variables when analysing data, and we intend to release the source code soon\footnote{The source code will be made available at http://galaxies-code.physics.ox.ac.uk.}. Whilst the GCS can be easily analysed in velocity space, angle-action variables should enable us to reveal structures, which are more dispersed in phase-space, in larger surveys.

We have limited the discussion in this paper to axisymmetric potentials. Whilst for our own spiral galaxy the axisymmetric approach may suffice, for analysing elliptical galaxies a triaxial approach must be developed. There are also triaxial St\"ackel potentials \citep[see][]{deZeeuw1985a} so it should be possible to expand the approach outlined in this paper to triaxial potentials. The extension of the angle-action estimation is simple, but the fitting procedure is more complex when the potential is triaxial. \citet{deZeeuw1985b} discuss how a general triaxial potential may be fitted both locally and globally by a St\"ackel potential. The method for global fitting is the three-dimensional generalisation of the method used in this paper so involves multiple multi-dimensional integrals. Also the best choice of coordinate system involves minimising the least-square difference with respect to two coordinate parameters so a more computationally expensive procedure than the simple method used in this paper may be required for finding the best coordinate system.

\section*{Acknowledgements}
I thank James Binney for carefully reading multiple drafts of this work and providing the code for calculating actions using the adiabatic approximation. 
I thank Paul McMillan for providing his torus-generating code and the code for calculating the potential of his best-fit mass model, and Andy Eyre for providing his code for calculating the actions and angles in an axisymmetric St\"ackel potential. I also thank the referee, Walter Dehnen, for many useful comments and I acknowledge the support of STFC.
\bibliographystyle{mn2e-2}
\bibliography{AnglesActionsInAxisymmetricPotential-JasonSanders}


\appendix
\section{Computing the angle-action variables}\label{AppendixAACalc}
The approach taken here as well as the majority of the formulae have been taken from \citet{Eyre2010}. Following on from equation \eqref{actionDef} the action $J_\tau$ is given by an integral over a full oscillation in $\tau$. A full oscillation in $\lambda$ involves integrating twice over the interval $(\lambda_0,\lambda_1)$. $\lambda_0$ and $\lambda_1$ are the roots of $p_\lambda$ which may be found by Brent's method using equation \eqref{TauEqOfMotion}. There is a complication when calculating $J_\nu$ due to the definition of $\nu$. $\nu$ is only uniquely defined for $z\geq0$ such that a full oscillation in $\nu$ corresponds to half an oscillation in $z$. Therefore we calculate $J_\nu$ by integrating four times over the interval $(\nu_0,\nu_1)$, where $\nu_0=c^2$, as all orbits cross the $z=0$ plane, and $\nu_1$ is the root of $p_\nu$ found by Brent's method. The actions are given explicitly by
\begin{equation}
J_\lambda = \frac{1}{\pi}\int_{\lambda_0}^{\lambda_1}p_\lambda \mathrm{d}\lambda,\quad
J_\nu = \frac{2}{\pi}\int_{\nu_0}^{\nu_1}p_\nu\mathrm{d}\nu.
\label{ExpicitAct}
\end{equation}

In order to calculate the angle coordinates we use the generating function, $S$, defined as
\begin{equation}
\begin{split}
S(\tau,\phi,E,L_z,I_3) &=  S_\phi + \sum_{\tau=\lambda,\nu}S_\tau,\\
&=\int_0^\phi L_z\mathrm{d}\phi' + \sum_{\tau=\lambda,\nu} \int_{\tau_0}^\tau p_{\tau'}\mathrm{d}\tau' .
\end{split}
\end{equation}
This generating function defines the canonical transformation between the canonical coordinates $(\tau, \phi, p_\tau, L_z)$ and $(J_\tau, L_z, \theta_\tau, \theta_\phi)$. 
The angles are now computed as
\begin{equation}
\theta_\tau = \frac{\partial S}{\partial J_\tau} = \frac{\partial S}{\partial E}\frac{\partial E}{\partial J_\tau}+\frac{\partial S}{\partial L_z}\frac{\partial L_z}{\partial J_\tau}+\frac{\partial S}{\partial I_3}\frac{\partial I_3}{\partial J_\tau},
\end{equation}
for $\tau=\lambda,\nu$.
The derivatives of the classical integrals with respect to the actions may be found by inverting the 3-by-3 matrix of the derivatives of the actions with respect to the classical integrals. These derivatives are simpler to calculate as they follow from equation \eqref{ExpicitAct} and the definition of $p_\tau$ from equation \eqref{TauEqOfMotion}:
\begin{equation}
\frac{\partial J_\lambda}{\partial E} = \frac{1}{4\pi}\int_{\lambda_0}^{\lambda_1} \frac{\mathrm{d}\lambda}{(\lambda-a^2)p_\lambda},
\end{equation}
\begin{equation}
\frac{\partial J_\lambda}{\partial L_z} = -\frac{L_z}{4\pi}\int_{\lambda_0}^{\lambda_1} \frac{\mathrm{d}\lambda}{(\lambda-a^2)^2p_\lambda},
\end{equation}
\begin{equation}
\frac{\partial J_\lambda}{\partial I_3} = -\frac{1}{4\pi}\int_{\lambda_0}^{\lambda_1}\frac{\mathrm{d}\lambda}{(\lambda-a^2)(\lambda-c^2)p_\lambda},
\end{equation}
\begin{equation}
\frac{\partial J_\nu}{\partial E} = \frac{1}{2\pi}\int_{\nu_0}^{\nu_1} \frac{\mathrm{d}\nu}{(\nu-a^2)p_\nu},
\end{equation}
\begin{equation}
\frac{\partial J_\nu}{\partial L_z} = -\frac{L_z}{2\pi}\int_{\nu_0}^{\nu_1} \frac{\mathrm{d}\nu}{(\nu-a^2)^2p_\nu},
\end{equation}
\begin{equation}
\frac{\partial J_\nu}{\partial I_3} = -\frac{1}{2\pi}\int_{\nu_0}^{\nu_1}\frac{\mathrm{d}\nu}{(\nu-a^2)(\nu-c^2)p_\nu}.
\end{equation}

The derivatives of the generating function with respect to the classical integrals may be calculated in the same spirit as
\begin{equation}
\frac{\partial S}{\partial E} = \sum_{\tau=\lambda,\nu}\frac{1}{4}\int_{\tau_0}^\tau\frac{\mathrm{d}\tau'}{(\tau'-a^2)p_{\tau'}},
\end{equation}
\begin{equation}
\frac{\partial S}{\partial L_z} = \phi-\sum_{\tau=\lambda,\nu}\frac{L_z}{4}\int_{\tau_0}^\tau\frac{\mathrm{d}\tau'}{(\tau'-a^2)^2p_{\tau'}},
\label{thetaphi}
\end{equation}
\begin{equation}
\frac{\partial S}{\partial I_3} = -\sum_{\tau=\lambda,\nu}\frac{1}{4}\int_{\tau_0}^\tau\frac{\mathrm{d}\tau'}{(\tau'-a^2)(\tau'-c^2)p_{\tau'}}.
\end{equation}
We note that equation \eqref{thetaphi} is simply the angle conjugate to $L_z$, $\theta_\phi$.

With the scheme given above there is a degeneracy in $\theta_\nu$ between points in the orbit at $\pm z$. This is simply resolved by adding $2\pi$ to $\theta_\nu$ if $z<0$. We then must divide $\theta_\nu$ by two to ensure $\theta_\nu$ is confined to the interval $(0,2\pi)$.

As $p^2(\tau)$ vanishes at the endpoints of many of these integrals, we want to avoid evaluating the integrands at the endpoints. We do this by performing a change of variables and estimating the integral using a Gauss-Legendre quadrature scheme. Here we will outline the procedure for calculating $J_\lambda$ but the same principle follows for the rest of the integrals. We perform a change of variables to
\begin{equation}
\lambda = \hat{\lambda}\sin\vartheta+\bar{\lambda};\>\bar{\lambda} = \frac{1}{2}(\lambda_0+\lambda_1);\>\hat{\lambda} = \frac{1}{2}(\lambda_1-\lambda_0),
\end{equation}
such that the integral is now over $\vartheta = (-\frac{\pi}{2},\frac{\pi}{2})$:
\begin{equation}
J_\lambda = \frac{1}{\pi}\int_{-\pi/2}^{\pi/2} \hat{\lambda}\cos\vartheta\,p(\lambda(\vartheta))\mathrm{d}\vartheta.
\end{equation}
This integral can now be computed numerically using a 10-point Gaussian-Legendre quadrature scheme.

\section{Derivation of Best-Fit Stackel Potential Functions}
\label{BestFitStackAppendix}
To find the best-fit St\"ackel potential we must minimise equation \eqref{leastsq} with respect to the function $f$. It is useful to consider minimisation with respect to the two parts of the function, $f(\lambda)$ and $f(\nu)$. This yields
\begin{equation}
\begin{split}
\int_{\lambda_-}^{\lambda_+}\mathrm{d}\lambda\,\Lambda(\lambda)[\chi(\lambda,\nu)-f(\lambda)+f(\nu)]&=0,\\
\int_{\nu_-}^{\nu_+}\mathrm{d}\nu\,N(\nu)[\chi(\lambda,\nu)-f(\lambda)+f(\nu)]&=0.
\end{split}
\end{equation}
Rearranging each of these, and noting that $\Lambda(\lambda)$ and $N(\nu)$ are normalized over the integration range, we find
\begin{equation}
\begin{split}
f(\lambda) &= \bar{\chi}(\lambda)+\int_{\nu_-}^{\nu_+}\mathrm{d}\nu\,N(\nu)f(\nu),\\
f(\nu) &= -\bar{\chi}(\nu)+\int_{\lambda_-}^{\lambda_+}\mathrm{d}\lambda\,\Lambda(\lambda)f(\lambda),
\end{split}
\label{rawf}
\end{equation}
where the definition of $\bar{\chi}(\tau)$ is given in equation \eqref{chibarFormulae}. Substitution of the expression for $f(\nu)$ into the expression for $f(\lambda)$ we find
\begin{equation}
f(\lambda) = \bar{\chi}(\lambda)-\bar{\bar{\chi}}+\int_{\lambda_-}^{\lambda_+}\mathrm{d}\lambda\,\Lambda(\lambda)f(\lambda),
\end{equation}
where $\bar{\bar{\chi}}$ is defined in equation \eqref{chibarFormulae}. This equation along with \ref{rawf} implies that
\begin{equation}
\int_{\lambda_-}^{\lambda_+}\mathrm{d}\lambda\,\Lambda(\lambda)f(\lambda)-\int_{\nu_-}^{\nu_+}\mathrm{d}\nu\,N(\nu)f(\nu) = \bar{\bar{\chi}}.
\end{equation}
As we are only constraining the difference $[f(\lambda)-f(\nu)]$ we are free to choose the values of these integrals as long as their difference equals $\bar{\bar{\chi}}$. We opt for the symmetric choice
\begin{equation}
f(\lambda) = \bar{\chi}(\lambda) - \frac{1}{2}\bar{\bar{\chi}},\>\> f(\nu) = -\bar{\chi}(\nu) + \frac{1}{2}\bar{\bar{\chi}}.
\end{equation}

\bsp

\label{lastpage}

\end{document}